\documentstyle[11pt,graphicx]{article}
\begin{document}
\title{Recent news and results in the computation 
       of NLO processes with 
new techniques\footnote{Contribution to the proceedings of the 10th Hellenic
        School on Elementary Particle Physics and Gravity, Corfu 2010.}}
\author{\sc{Roberto Pittau}\\
{\it Departamento de F\'{i}sica Te\'orica y del Cosmos and CAFPE} \\
{\it U. de Granada, E-18071 Granada, Spain and} \\ 
{\it PH Department, TH Unit, CERN, CH-1211 Geneva 23, Switzerland}}
\renewcommand{\today}{}
\maketitle

\begin{abstract}
  I illustrate new techniques and recently obtained results in the computation of perturbative QCD processes at the NLO accuracy.
\end{abstract}

\section{Introduction}
 The field radiative corrections for multi-particle 
processes has received a lot of attention, in the last few years, 
also thanks to new computational 
techniques~\cite{Ossola:2006us,Ellis:2009zw,Berger:2009ep}
 and to new tools~\cite{Ossola:2007ax,Giele:2008bc,vanHameren:2009dr,Hirschi:2011pa} 
that are nowadays available. 
Next to Leading Order (NLO) QCD calculations for Hadron Colliders physics  
are needed for two reasons.
Firstly, they are an important ingredient for reliably computing 
backgrounds in new physics searches, that very often rely on analyses
performed in rather narrow corners of the Phase-Space or 
in tails of distributions, where, on the one hand, not enough statistics is present to extract
the background directly from the data and where, on the other hand, radiative 
corrections are expected to be large. Secondly, NLO calculations should be always preferred when measuring and constraining fundamental quantities of the Standard Model (SM), such as the mass of the Higgs particle (if found)
and its couplings, $M_W$, $\alpha_S$ or $M_{\rm top}$.
Both in the case of new physics searches, where the new produced particles 
undergo long decay chains, and in the case of SM measurements, 
where the hard event is accompanied by a rather strong jet activity,
multi-leg final states are expected as a typical signature.

 In this contribution, I review the main results that have been recently 
obtained in the subject of NLO QCD calculation for multi-leg final states
observables.

\section{NLO processes needed at the LHC}
In Les Houches 2007, theoreticians and experimentalists agreed upon a list 
of processes interesting to know at the NLO accuracy in QCD~\cite{Bern:2008ef}. In occasion of the following NLO multi-leg Les Houches workshop~\cite{Binoth:2010ra}, the job was already almost accomplished, at least at the parton level, mainly due to new breakthrough techniques to compute the 
one-loop part of the NLO corrections~\cite{Ossola:2006us,Ellis:2009zw,Berger:2009ep}.

For reader's reference I present, in table~\ref{tab:tab1}, 
the original list and the few entries added in 2009. 
\begin{table}
\begin{center}
\begin{tabular}{|lll|}
\hline 
$pp \to W+j$ &$ pp \to t \bar t + 2j$&$ pp \to V + 3j$ \\
$ pp \to H+2j$&$ pp \to VVb \bar b$   &$ {pp \to t \bar t b\bar b}$ \\
$ pp \to VVV$ &$ pp \to VV + 2j$      &$ pp \to b \bar b b \bar b$ \\
\hline 
\hline 
$pp \to t \bar t t \bar t $ &$ pp \to 4j$&$ pp \to W + 4j$ \\
$ pp \to Z+3j$&$ pp \to W b \bar b j$   & \\ 
    \hline
\end{tabular}
\end{center}
  \caption{The original 2007 Les Houches Wish List (top) 
and its 2009 update (bottom).}
  \label{tab:tab1}
\end{table}
At present, all the parton level processes in table~\ref{tab:tab1}, 
except $pp \to 4j$, are known at the the NLO accuracy in QCD, in the sense
that theory papers have been written containing NLO results and distributions.
However, it has to be pointed out that, even if table \ref{tab:tab1} looks 
quite impressive, the final NLO product, needed from 
an experimental point of view, should be a usable code, fully automatic and 
matched with Parton Shower and Hadronization.

Progress in the direction of a complete full automation of parton level
NLO predictions has been achieved very recently by the authors of the
{\sc MadLoop} code of~\cite{Hirschi:2011pa}, where the ability 
of {\sc MadGraph} ~\cite{Maltoni:2002qb} to compute amplitudes 
is merged with the OPP 
integrand reduction method implemented in {\sc CutTools} 
to automatically generate one-loop corrections. On the other hand, interfacing with the Parton Shower and
Hadronization is possible within the MC@NLO~\cite{Frixione:2002ik} and 
POWHEG~\cite{Frixione:2007vw} frameworks. 

Finally, the first attempt of automatizing {\em both} the NLO computations 
{\em and} the subsequent merging with the Parton Shower and Hadronization
codes is under way, under the name a{\sc MC@NLO}~\cite{Frederix:2011zi,ref2}.

\section{NLO Tools}
It is evident that sophisticated programs are needed to compute multi-leg 
processes at NLO. The existing tools 
can be naturally divided in three categories, as listed in table 
\ref{tab:tab2},
namely codes based on Analytic Formulae, on traditional Feynman Diagram
techniques and, finally, on OPP or Generalized Unitarity methods.
\begin{table}
\begin{center}
\begin{tabular}{|l|}
\hline\\
{\underline{Analytic Formulae:}} \\ 
\\
{\sc MCFM}~\cite{Campbell:2002tg} \\\\
{\underline{Feynman Diagrams:}}  \\ 
\\
{\sc Bredenstein, Denner, Dittmaier, Pozzorini}~\cite{Bredenstein:2009aj} \\
{\sc FormCalc/LoopTools/FeynCalc}~\cite{Hahn:1998yk} \\
{\sc Golem}~\cite{Binoth:2010pb} \\\\
{\underline{OPP/Generalized Unitarity:}}  \\
\\{\sc MadLoop}~\cite{Hirschi:2011pa} \\
{\sc Helac-NLO/Cuttools}~\cite{Ossola:2007ax,vanHameren:2009dr}
 \\
{\sc BlackHat/Sherpa}~\cite{Berger:2009ep} \\
{\sc Rocket}~\cite{Giele:2008bc} \\
{\sc Golem/Samurai}~\cite{Mastrolia:2010nb,Heinrich:2010ax}\\ 
\hline
\end{tabular}
\caption{Some available NLO tools. \label{tab:tab2}}
  \end{center}
\end{table}
As usual, most of the programs have been cross checked, to establish their
technical agreement. An example of such {\em tuned} comparisons 
is reported in table \ref{tab:tab3}, for the process $pp \to ttbb$.
It is a remarkable fact that the two codes use two completely 
different techniques.
\begin{table}
\begin{center}
\begin{tabular}{|c | c c | c c |}
\hline
&&&& \\
Process & 
$\sigma^{\mbox{\footnotesize{LO}}}$    [fb]  {\cite{Bredenstein:2009aj}}     
& $\sigma^{\mbox{\footnotesize{LO}}}$  [fb]  {\cite{vanHameren:2009dr}}   
& $\sigma^{\mbox{\footnotesize{NLO}}}$ [fb]  {\cite{Bredenstein:2009aj}}  
& $\sigma^{\mbox{\footnotesize{NLO}}}$ [fb]  {\cite{vanHameren:2009dr}}  
  \\
&&& & \\
\hline
&&&&\\
$ q \bar{q}\rightarrow t\bar{t}b\bar{b} $  
& 85.522(26) 
& 85.489(46)
& 87.698(56)
& 87.545(91) 
   \\
&&&&\\
\hline 
&&&&\\
$ pp\rightarrow t\bar{t}b\bar{b} $  
& 1488.8(1.2)  
& 1489.2(0.9)
& 2638(6) 
& 2642(3)
 \\
&&&&\\
\hline
\end{tabular}
\end{center}
\caption{Example of  {\em tuned} comparisons between 
HELAC-NLO~\cite{vanHameren:2009dr} and the 
program of~\cite{Bredenstein:2009aj}. \label{tab:tab3}}
\end{table}
Analogous successful comparisons have been performed by the {\sc Golem} group and
the team Dittmaier, Kallweit and Uwer on $pp \to ZZ+j + X$~\cite{Binoth:2010ra}.

\begin{figure}
\begin{center}
\includegraphics[scale=0.9]{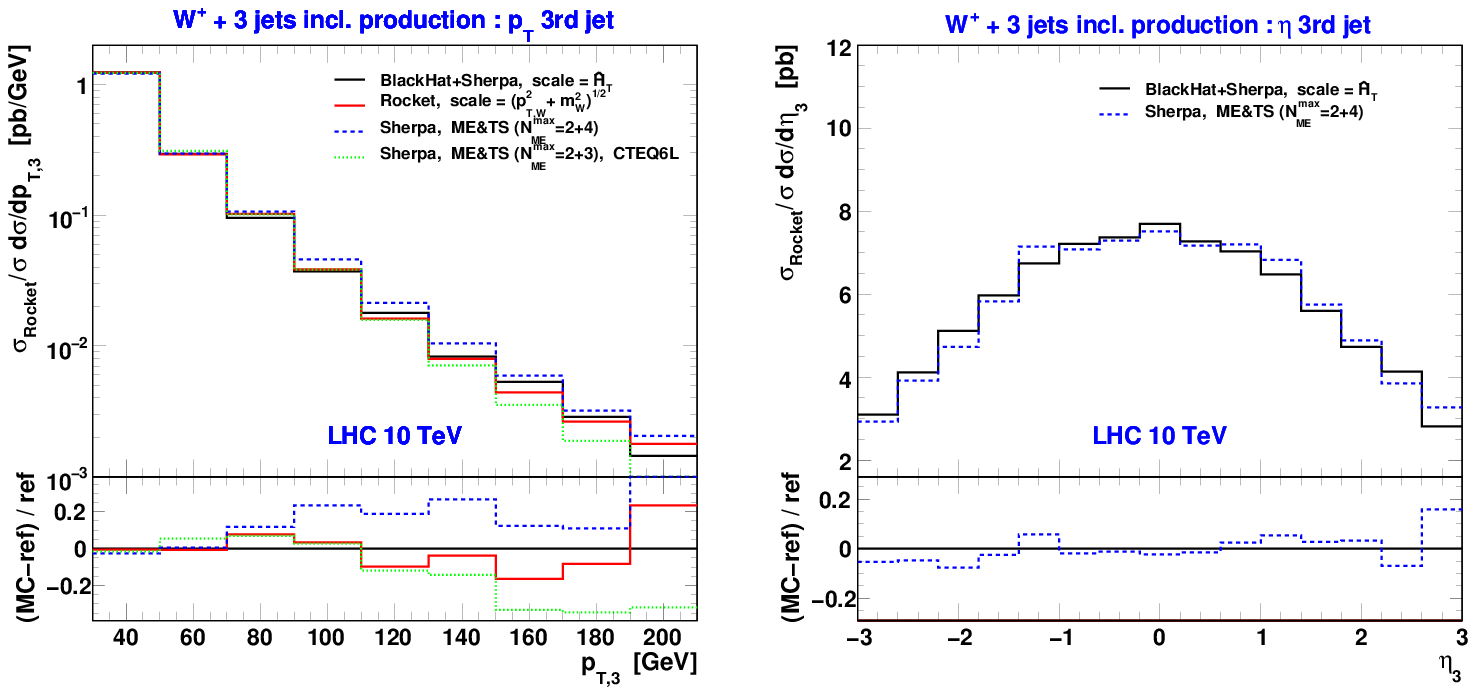}.
\end{center}
  \caption{Comparisons on $pp \to W+ 3jets$: $p_t$ and 
rapidity of the 3rd jet.   \label{fig:fig1}}
\end{figure}

The second, even more important task of the comparisons, 
is the assessment of the
theoretical accuracy at which a given process is known. In this second type of
exercise, each program freely varies a few parameters 
(such as renormalization and factorization scales). The goodness of the LO
prediction (at least in the shape of the distributions) can also be
determined that way.  In fig.~\ref{fig:fig1} I report, as an example, 
the result of a comparison of {\sc BlackHat}, {\sc Rocket} and {\sc Sherpa}
on   $pp \to W+ 3jets$ at NLO.

\begin{figure}
\begin{center}
\includegraphics[scale=1]{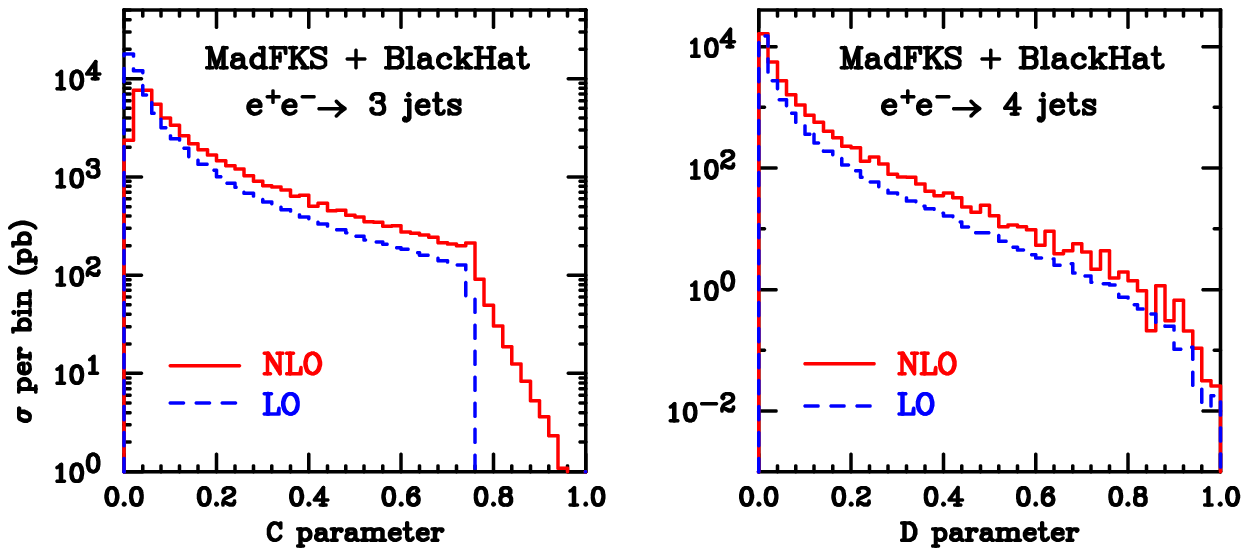}.
\caption{Results on $e^+e^- \to jets$ using the Binoth Les Houches accord.
\label{fig:fig2}}
\end{center}
\end{figure}

As it can be easily understood, with the advent of the LHC data the techniques used to obtain the NLO 
results are getting less and less important, since the interest is now going towards commonly 
accepted interfaces to merge different parts of the NLO calculations. 
As an example, an accord to interface Monte Carlo (MC) programs, generating 
the real radiation, together with programs providing the 
virtual one-loop contributions (OLP), can be found in~\cite{Binoth:2010xt}
(the so called Binoth Les-Houches accord).
In fig.~\ref{fig:fig2},  I show this accord at work between 
{\sc BlackHat/Rocket} on the
OLP side and {\sc MadFKS}~\cite{Frederix:2009yq} 
on the MC side, in the case of  $e^+e^- \to jets$ 
as implemented by Frederix, Maitre and Zanderighi~\cite{Binoth:2010ra}.
The Binoth Les-Houches accord is also used by 
a{\sc MC@NLO}~\cite{Frederix:2011zi,ref2} to interface virtual and real corrections.
  
Finally, it should be noticed that the field is evolving so rapidly that
new NLO processes are continuously computed. As an illustrative example, I quote
$pp \to W^+W^\pm jj$ in~\cite{Melia:2010bm,Melia:2011dw}, $pp \to t \bar t \to W^+W^-b \bar b$ including
all off-shell effects in~\cite{Denner:2010jp,Bevilacqua:2010qb} 
and $pp \to Wjjjj$ ~\cite{Berger:2010zx}.

\section{Conclusions}
I have presented recent progresses in our theoretical understanding
of perturbative QCD at NLO. New automatic NLO tools exist nowadays to deal with 
the LHC data and to cope with the complexity of the present and forthcoming
measurements. The further step of automatically interfacing such programs 
with Parton Shower and Hadronization has already been undertaken.

\section*{Acknowledgment}
Work supported by the Spanish MEC under project FPA2008-02984.

\end{document}